\documentclass[twoside,12pt]{article}  %%% 这是 LaTeX 格式文件必须的指令，不可缺少
\usepackage{CJK}   %%%%% 使用中文、日文、韩文时调用此宏包
\usepackage{indentfirst}  %% 节标题后第一行段缩进
\usepackage{bm}    %%%%%  排印数学符号黑斜体时调用此宏包
\usepackage{float}
\usepackage{graphicx}  %%%% 文中插入 eps 图形时调用此宏包，插图方法 1
\begin{document}    %% 文本文件开始，这是必须的指令

\begin{CJK*}{GBK}{song}  %% 开始进入中文环境

%-------------------  First Head  -----------------------------------------
\thispagestyle{empty} \vspace*{0.8cm}\hbox
to\textwidth{\vbox{\hfill\huge\sf \hfill}}
%to\textwidth{\vbox{\hfill\huge\sf Chinese Physics B\hfill}}
\par\noindent\rule[3mm]{\textwidth}{0.2pt}\hspace*{-\textwidth}\noindent
\rule[2.5mm]{\textwidth}{0.2pt}

%=================== Text begin here =============================================

\begin{center}
\LARGE\bf Simple rate-adaptive LDPC coding for quantum key distribution   %% 论文题目
\end{center}

\footnotetext{\hspace*{-.45cm}\footnotesize $^*$Project supported by the National Basic Research Program of China (2011CBA00200 and 2011CB921200), the National Natural Science Foundation of China (61101137, 61201239, 61205118, and 11304397), and China Postdoctoral Science Foundation (2013M540514).}
\footnotetext{\hspace*{-.45cm}\footnotesize Corresponding authors. E-mail: ${}^{\star}$yinzheqi@mail.ustc.edu.cn; ${}^{\dagger}$zfhan@ustc.edu.cn. }

\begin{center}
\rm Mo Li$^{\rm 1,2}$, \ \ Chun-Mei Zhang$^{\rm 1,2}$, \ \ Zhen-Qiang Yin$^{\rm 1,2\star}$,\ \ Wei Chen$^{\rm 1,2}$,\ \ Chuan Wang$^{\rm 1,2}$, \ and  \ Zheng-Fu Han$^{\rm 1,2\dagger}$
\end{center}

\begin{center}
\begin{footnotesize} \sl
${}^{\rm 1}$    Key Laboratory of Quantum Information, CAS, University of Science and Technology of China, Hefei 230026, China\\
${}^{\rm 2}$    Synergetic Innovation Center of Quantum Information \& Quantum Physics, University of Science and Technology of China, Hefei 230026, China\\

\end{footnotesize}
\end{center}

%\begin{center}
%\footnotesize (Received X XX XXXX; revised manuscript received X XX XXXX)
%          %% (Received 日 月 年; revised manuscript received 日 月 年)
%\end{center}

\vspace*{2mm}

\begin{center}
\begin{minipage}{15.5cm}
\parindent 20pt\footnotesize
Although quantum key distribution (QKD) comes from the development of quantum theory, the implementation of a practical QKD system does involve a lot of classical process, such as key reconciliation and privacy amplification, which is called post-processing. Post-processing has been a crucial element to high speed QKD systems, even the bottleneck of it because of its relatively high time consumption. Low density parity check (LDPC) is now becoming a promising approach of overcoming the bottleneck due to its good performance in processing throughput. In this article we propose and simulate an easily implemented but efficiently rate-adaptive LDPC coding approach of reconciliation, different from the previously proposed puncturing- and shortening-based approach. We also give a measure for choosing the optimal LDPC parameter for our rate-adaptive approach according to error rates.
\end{minipage}
\end{center}

\begin{center}
\begin{minipage}{15.5cm}
\begin{minipage}[t]{2.3cm}{\bf Keywords:}\end{minipage}
\begin{minipage}[t]{13.1cm}
LDPC, rate-adaptive, throughput, quantum key distribution (QKD).
\end{minipage}\par\vglue8pt
{\bf PACS: }
%%% PACS 分类码
%% 查询网址：http://www.aip.org/pacs
\end{minipage}
\end{center}

\section{Introduction}  %%% 节标题 1
Quantum key distribution (QKD) \cite{BB84} allows two communication parties Alice and Bob to share unconditional secret keys, in the presence of an eavesdropper Eve. Generally, QKD contains quantum communication phase and post-processing phase \cite{Cascade,Winnow,LDPC}. Quantum theory states that the two key strings will be perfectly the same with each other. Unfortunately, due to the disturbance of environment and the imperfection of devices, there are in fact some discrepancies between the two key strings, keeping it from being utilized directly. Thus, in order to make their key strings same with each other, Bob should correct his \lq\lq wrong bits\rq\rq according to Alice's string, or vice versa, which is called reconciliation.

There have been several schemes proposed for reconciliation, such as Cascade \cite{Cascade}, Winnow \cite{Winnow} and LDPC \cite{LDPC}. Cascade and Winnow both can correct the errors with high efficiency. However, they demand many times of interactive communication between Alice and Bob, while in high-speed QKD systems \cite{Wang,Dixon,Akihiro Tanaka} communication times have to be harshly restricted for saving time. Different from the above two schemes, LDPC code can get rid of interactive communication and largely reduce the communication time. Furthermore, the encoding and decoding algorithm of LDPC code in QKD scenario slightly varies from that in classical scenario \cite{LDPC} and is more easily realized and favorable to high throughput. LDPC encoding in QKD system is commonly referred as syndrome encoding and the transmitted messages are merely the redundant bits. It needs only the classical sparse $H$ matrix from LDPC code in both encoding and decoding phase, which can specially benefit the processing speed. What's more, due to its parallel-like encoding and decoding algorithm, LDPC can be implemented on GPU, which can dramatically boost the processing speed.
So far there have been many discussions on LDPC implementation in QKD system and some approaches to improve LDPC's performance in QKD situation, such as puncturing, shortening, and blind reconciliation \cite{David Elkouss, Martinez-Mateo, David Elkouss 2}. In most discussions, the reconciliation efficiency is the mainly studied object, which in syndrome encoding can be written as
\begin{equation}
f = \frac{{1 - r}}{{h(e)}}
\end{equation}
where $r$ is the rate of the matrix used in reconciliation, and ${h(e)}$ is the Shannon entropy. The efficiency indicates how far the disclosure in the reconciliation protocol is from the minimum theoretical information disclosure. If the rate is variable during the reconciliation, the overall efficiency is the averaged value
\begin{equation}
f = \frac{{\mathop \sum \nolimits_{i} \frac{{1 - {r_i}}}{{h(e)}}{n_i}}}{{\mathop \sum \nolimits_i {n_i}}}
\end{equation}
where ${{n_i}}$ refers to the adoption times of the rate ${{r_i}}$. Although reconciliation efficiency is an important index of performance, the real strength of LDPC code, as mentioned in the abstract, lies more in its high throughput capability and few interactive communication rounds. So the final throughput should also be well investigated in order to better utilize LDPC in QKD systems. There are much fewer work \cite{Jesus Martinez-Mateo} considering the throughput performance of LDPC code compared to those on efficiency.

Here, we propose an easy but very efficient method to build a rate-adaptive reconciliation approach and give a measure for choosing suitable parameters for optimization. Our rate-adaptive approach has only one adjusting rule, which is different from those based on puncturing and shortening technique \cite{David Elkouss 2}.

The article is organized as follows. First we elaborate our rate-adaptive approach and give the measure for optimizing its performance. Then a simulation is given using the optimal method.

\section{EASY RATE-ADAPTIVE LDPC CODE}  %%% 节标题 2
\begin{figure}[H]
\centering
\includegraphics[width=0.4\textwidth]{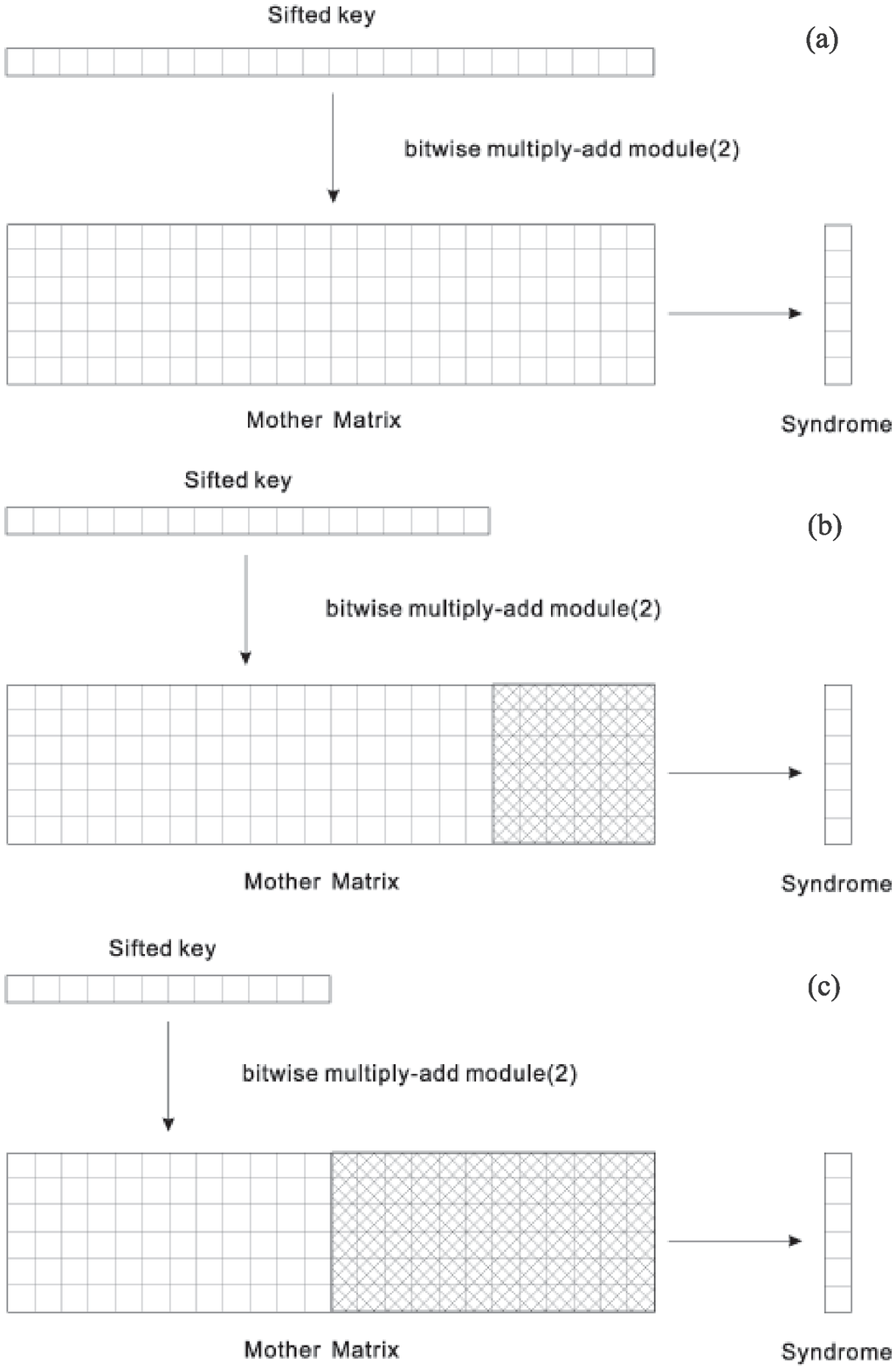}
\parbox{15.5cm}{\small{\bf Fig.1.}  $a$) Each bit of syndrome is generated by a multiply-add module $2$ operation of sifted key and the corresponding row of encoding matrix ($H$ matrix). $b$) The sifted key is changed to $3/4$ the width of $H$ matrix. $c$) The sifted key is changed to half the width of $H$ matrix. The relation of error rates that the above three situations adapt to is : ${e_c} > {e_b} > {e_a}$. }
\label{figure1}
\end{figure}

In the application of LDPC code in QKD reconciliation, only redundant information is sent, which is referred as syndrome of sifted keys. It is generated by module $2$ bitwise multiply-add operation between the sifted key bits and rows of the parity check matrix, as shown in Fig. \ref{figure1} $a$). Usually the matrix is fixed because the change of matrix consumes huge resource, such as time and memory space, and slows the throughput. But fixing the matrix also means that the information disclosed during reconciliation is fixed regardless of the error rate level. This will obviously lead to a waste of the secure information substantially when the error rate is tiny, resulting in a bad efficiency.

In this article, with a fixed mother matrix, we adaptively adjust the length of encoding sifted keys according to the error rate level and achieve a good balance between throughput and efficiency. The strategy is stated as follows. We keep the syndrome string's length fixed. When the error rate is high, less sifted keys are encoded at a time, and more sifted keys are encoded when the error rate is low, as shown in Fig.1 $a$), $b$), and $c$). Hence, we only use certain part of the mother $H$ matrix in one encoding round depending on the error level. The used part can be called the effective matrix. And for better performance, we can in advance investigate the mother matrix and build a look-up table which can help choose a suitable effective matrix according to the error rate.

Assume that the size of parity check matrix is $m \times n$, i.e. the mother matrix, where $m$ is the height, i.e. syndrome string's length, and $n$ is the width, i.e. the encoding key string's length when the whole matrix is used. It is easily known that the code rate based on the mother matrix is
\begin{equation}
{r_0} = 1 - \frac{m}{n}
\end{equation}

However, since most of the time we only use the effective matrix, not the whole matrix, the true encoding rate is not always consistent with ${r_0}$. We can name the rate of the effective matrix as \emph{encoding efficient rate} ($EER$). For example, in Fig.1 $a$), $b$), and $c$), the $EER$s are ${r_0}$, ${r_0}/2$, and $3{r_0}/4$ respectively.

Obviously, as mentioned above, the computational complexity can be varied in this LDPC application. When only a part of the matrix is used, the computation complexity is lower than that when the whole matrix is used. This can decrease the average computational complexity of the reconciliation.

In order to change the encoding length of sifted keys efficiently and adaptively, it should be guaranteed that every used part of $H$ matrix, $m \times n'$, has good quality in girth, which is inextricably related to the capability of its error correction. A girth of $4$ will be bad and not very capable of correcting errors. This job can be perfectly accomplished by introducing the progressive edge growth(PEG) method \cite{Xu FangXing}. The matrix should be generated using PEG in the sequence of column by column but not row by row. We give the minimum girth with respect to the length of effective matrix in Fig.2 The mother matrix is $1024 \times 5120$
 (${\rm{height  \times  width}}$), with half of the symbol nodes being $4$-degree nodes and half being $5$-degree nodes.
\begin{figure}[H]
\centering
\includegraphics[width=\linewidth]{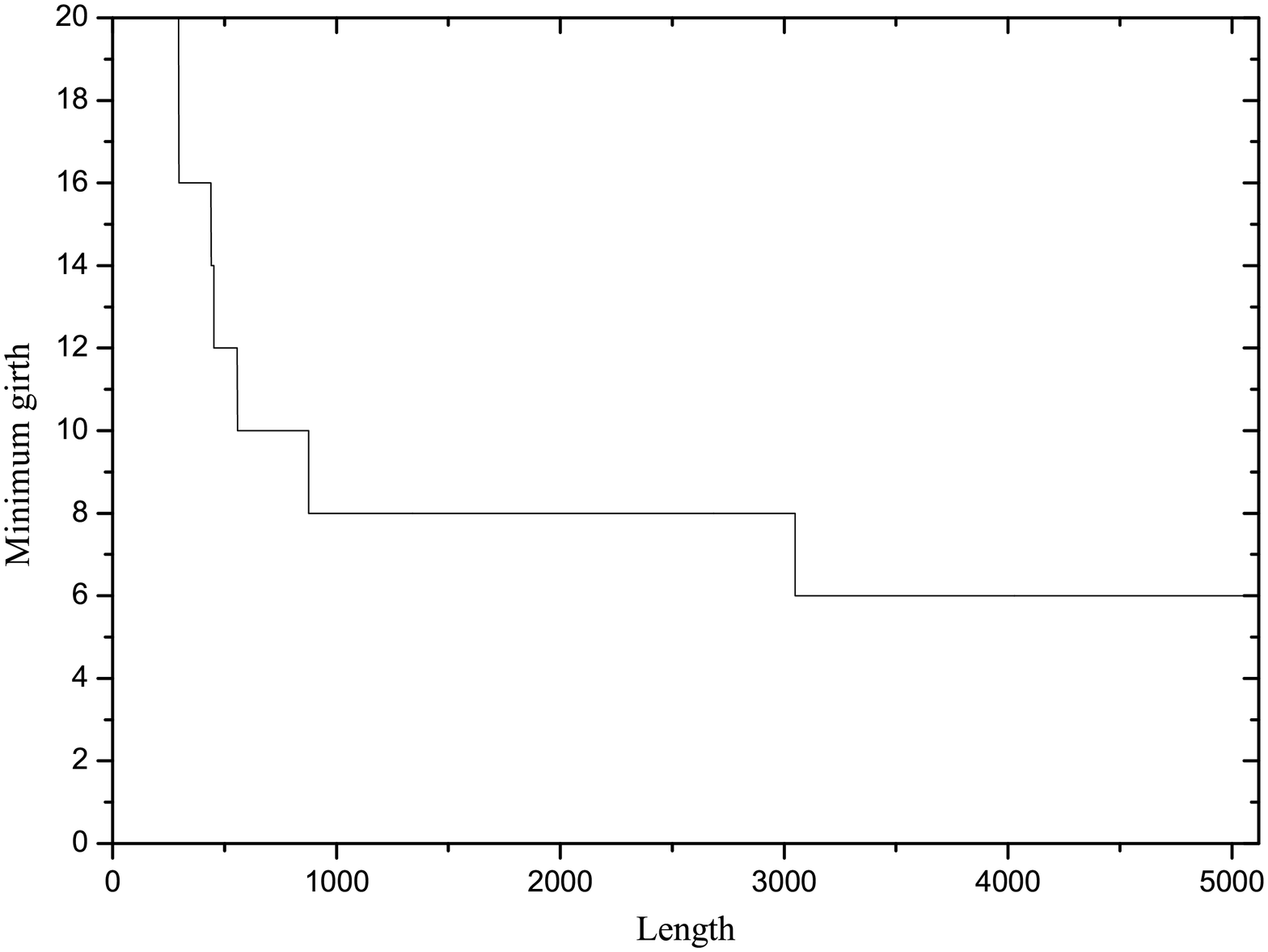}
\parbox{15.5cm}{\small{\bf Fig.2.}  The minimum girth with respect to the used part of the matrix that we will use in the simulation. We can see that when a small part of the matrix is used, the girth will be very large, and thus the matrix at this point has good error correcting capability. Even when it's 5120, the girth is still 6.}
\label{figure2}
\end{figure}

\section{CHOICE OF OPTIMAL LDPC CODE}
It is obvious that the secure key throughput will decline when the EER is decreased, because of more disclosure of information. So we should make it economically optimal that where the EER should be changed according to the error rate.

In a successful reconciliation, if the privacy amplification is not considered, the ratio of secure information distilled is equal to EER. The probability of conducting a successful reconciliation is ${(1 - FER)}$, where $FER$ is the frame error rate. Hence, the average ratio of secure information that can be distilled after a reconciliation is
\begin{equation}
\alpha  = (1 - FER) \times EER
\end{equation}
We call $\alpha $ the distillation efficiency. It can be seen that $\alpha $  is the property of the algorithm and independent of the processing device's capability. Now, for getting maximum amount of secure information, our goal is to find the maximum distillation efficiency given the error rate. We use the $1024 \times 5120$ matrix generated in the previous section as the mother matrix. Table 1 shows the value of $\alpha $  with respect to different lengths of effective matrix and different error rates.

 %%%%% 表格插入形式：
\begin{center}
\tabcolsep=15pt  %%% 此参数用于调整表的整体宽度
\small
\renewcommand\arraystretch{0.6}  %% 调整表内行与行之间的纵向距离
\begin{minipage}{15.5cm}{
\small{\bf Table 1.} Distillation Efficiency}
\end{minipage}
\vglue2pt
\begin{tabular}{c|cccc}
\hline %% 表横线
 Error rate(\%) &5120 &4096 &3072 \\     %% 第1行
 \hline
  1.0 &\bf80.00 &75.00 &66.67\\   %% 第2行
  1.1 &\bf80.00 &75.00 &66.67\\   %% 第3行  %%%% 依次往下排其他行
  1.2 &\bf80.00 &75.00 &66.67\\
  1.3 &\bf79.52 &75.00 &66.67\\
  1.4 &\bf76.32 &75.00 &66.67\\
  1.5 &64.48 &\bf75.00 &66.67\\
  1.6 &32.88 &\bf75.00 &66.67\\
  1.7 &-- &\bf75.00 &66.67\\
  1.8 &-- &\bf74.77 &66.67\\
  1.9 &-- &\bf73.80 &66.67\\
  2.0 &-- &48.30 &\bf66.67\\
  2.1 &-- &-- &\bf66.67\\
  2.2 &-- &-- &\bf66.67\\
  2.3 &-- &-- &\bf66.67\\
  2.4 &-- &-- &\bf66.67\\
  2.5 &-- &-- &\bf66.67\\
  2.6 &-- &-- &\bf65.60\\
  2.8 &-- &-- &\bf64.40\\
  3.0 &-- &-- &\bf60.47\\
\hline
\end{tabular}
\end{center}

In the table, $5120$ means the whole mother matrix is used, while $3072$ means only $2/3$ of the mother matrix from the left side is used (as shown in Fig.1.). In the table we don't list the data when $\alpha $ is too small. What's more, we don't list the high-error-rate situation, which is higher than ${\rm{3\% }}$, because our system runs typically under an error rate around ${\rm{1\% }}$ to ${\rm{2\% }}$. With such a $\alpha $ table, we can easily determine which encoding length we should choose in order to obtain a maximum distillation efficiency. Following this principle, the bold elements in Table 1 are the working region of our LDPC approach, which are also the optimal parameters for our simulation in the following section.

It should be particularly noted here that we assume that the processing speed of the device involved in reconciliation is highly sufficient for dealing with the generation of sifted keys, so that we can just focus on the choice of $\alpha $. The assumption is reasonable because now with a GPU, a speed of as high as a few dozens of Mbps can be realized \cite{David Elkouss 2}, which far outnumbers the speed of sifted key's generation in nowadays QKD system.

\section{SIMULATION}
We use our phase-coding decoy QKD system's physical parameter in the simulation. That is, the attenuation of fiber is 0.2dB/km, and the frequency of light source is 200MHz. The detector's efficiency is $0.1$, with a dark count probability of ${10^{ - 5}}$. The interference visibility is $0.98$. A light pulse contains average $0.6$ photon. Our LDPC protocol is one-way communication protocol, which means Bob only tells Alice if he successfully corrects his key. If the reconciliation fails, they abandon the related sifted keys and conduct the next reconciliation round. In the simulation the mother matrix's size is $1024 \times 5120$, and there are four choices of effective matrices, whose lengths are $5120$, $4096$, $3072$, and $2048$, respectively.

Fig.3 shows the result of our simulation. The secure key ratio is equal to the distillation efficiency introduced above. Here we don't take privacy amplification into account. It can be seen that as the distance increases the ratio keeps decreasing, which indicates smaller and smaller portion of secure information is distilled out of the sifted key because of more disclosure needed for error correction. The secure key's throughput is drawn compared to the sifted key's throughput in the figure. The bigger and bigger gap between these two lines are due to the decreasing secure ratio. At a short distance($20$km), errors can be corrected with just around ${\rm{20\% }}$ information disclosed, resulting in around ${\rm{80\% }}$ secure information preserved, while at a very long distance, such as $110$km, errors cannot be corrected even with the smallest effective matrix. Of course this can be improved by generating a more robust matrix against the error rate.

\begin{figure}[H]
\centering
\includegraphics[width=\linewidth]{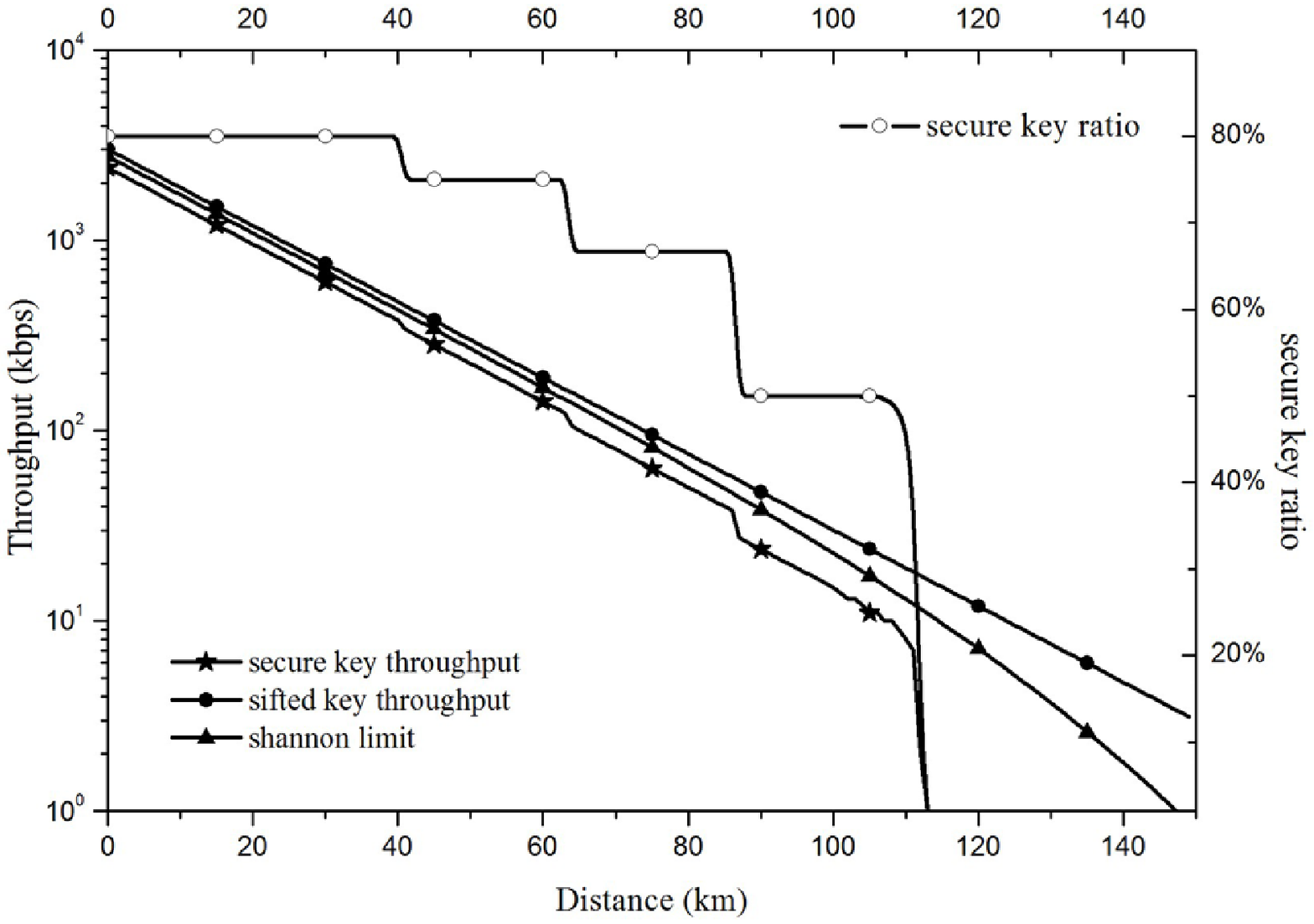}
\parbox{15.5cm}{\small{\bf Fig.3.} The result of simulation. The secure key ratio is the ratio between distill key amount and sifted key amount. Privacy amplification is not considered here. The obvious ladder-like behavior of secure key ratio is due to the discrete choice of the length of encoding sifted key and if the length is distributed intensely the cure will be smooth. The sudden decrease in the secure key throughput line results from the steep decline of secure key ratio.}
\label{figure3}
\end{figure}

\section{CONCLUSION}
We propose a simple rate-adaptive LDPC-based reconciliation method and simulate its performance with practical a QKD system. The result shows that our approach can change the rate according to different error rates and give an optimal secure keys throughput.

We also give a measure for configuring LDPC for the objective of optimal throughput. Due to its independence on processing devices, the measure is helpful on configuring LDPC in different QKD systems.

\vspace*{2mm}

%%%% 参考文献排版格式：

\end{CJK*}  %% 结束中文、日文、韩文使用环境
\end{document}